\documentstyle[12pt,english]{report}
\newenvironment{eqn}
{\begin{equation}\begin{array}}{\end{array}\end{equation}{}} 
 
\def\({\Bigl(} \def\){\Bigr)}\def\|{\Big|}
\def\o{\circ}\def\x{\times}
\def\ox{\otimes}
\def\pl{\oplus}
\def\SUM{\displaystyle \sum}
\def\mid{\big\bracevert}
\def\then{~\Rightarrow~}\def\subnoteq{\subset}
\def\and{\wedge}
\def\od{\vee}
\def\m{\bullet}

\def\A{{\,{\rm A\kern-.55emA}}}\def\C{{\,{\rm I\kern-.55emC}}}
\def\E{{\,{\rm I\kern-.2emE}}}\def\H{{\,{\rm I\kern-.2emH}}}
\def\I{{\,{\rm I\kern-.2emI}}}\def\K{{\,{\rm I\kern-.2emK}}}
\def\L{{\,{\rm I\kern-.2emL}}}\def\M{{\,{\rm I\kern-.16emM}}}
\def\N{{\,{\rm I\kern-.16emN}}}\def\Q{{\,{\rm I\kern-.5emQ}}}
\def\R{{\,{\rm I\kern-.2emR}}}\def\S{{\,{\rm I\kern-.42emS}}}
\def\T{{\,{\rm I\kern-.37emT}}}\def\Z{{\,{\rm Z\kern-.35emZ}}}
\def\rin{{\,\in\kern-.42em\in}}
 
\def\tr{{\,{\rm tr }\,}}

\def\FIX{\,\hbox{FIX}}

\def\p{\partial}
 
\def\al{\alpha}  \def\be{\beta} 
\def\de{\delta}  \def\ep{\epsilon}  \def\ze{\zeta}
\def\th{\theta}    
\def\ka{\kappa}   \def\la{\lambda}   \def\si{\sigma}
   \def\om{\omega} 
\def\phi{\varphi}    \def\Th{\Theta}
    \def\La{\Lambda}

\let\rvec=\vec        
\def\vec#1{\underline{\bf vec}_{#1}}

 \def\GL{{\bf GL}}  \def\SL{{\bf SL}}
\def\U{{\bf U}} \def\O{{\bf O}}   \def\SU{{\bf SU}} \def\SO{{\bf SO}}

\def\d#1{\check{#1}}\def\angle#1{\langle#1\rangle}

\def\brack#1{\lbrack#1\rbrack}\def\brace#1{\lbrace#1\rbrace}

\def\ul#1{\underline{#1}}
\def\ol#1{\overline{#1}}\def\bl#1{{\bf #1}}\def\cl#1{{\cal #1}}
\def\ro#1{{\rm #1}}

\def\dprod#1#2{\langle#1,#2\rangle}
\def\sprod#1#2{\langle#1|#2\rangle}
\def\com#1#2{\lbrack#1,#2\rbrack}

\def\acom#1#2{\{#1,#2\}}

\def\map{\longrightarrow}

\def\mape{\longmapsto}

\begin{document}
\begin{titlepage}

\hfill MPI-PhT/96-58 (July 96)
\vskip3cm
\centerline{\bf THE COUPLING CONSTANTS AND MASSES}
\centerline{\bf OF THE STANDARD MODEL}
\centerline{\bf AS SYMMETRY NORMALIZATIONS}
\vskip2cm
\centerline{Heinrich Saller\footnote{\scriptsize 
e-mail adress: hns@mppmu.mpg.de}}
\centerline{Max-Planck-Institut f\"ur Physik und Astrophysik}
\centerline{Werner-Heisenberg-Institut f\"ur Physik}
\centerline{M\"unchen}
\vskip25mm

\centerline{\bf Abstract}

The numerical input for the  quantitative consequences 
of the electroweak standard model,
the hypercharge and isospin 
coupling constants and the Higgs field ground state mass value,
are interpreted as normalizations of the symmetries involved.
Using an additional statistical argument, a
first order quantitative determination of the Weinberg angle 
as a normalization ratio gives the experimentally acceptable 
value $\tan^2\th={1\over3}$.

\end{titlepage}

\advance\topmargin by -1.6cm

\setcounter{page}{1}
$~~$\vskip2cm
\centerline{\bf Introduction}
\vskip1cm
The standard model of the electroweak interactions\cite{WEIN} derives
its quantitative predictions with the experimental
input of two independent  coupling constants $g_1,g_2$ for the hypercharge 
$\U(1)$ and isospin $\SU(2)$
gauge fields resp. and the 
harmonic particle analysis fixing ground state value 
$M$ of the 
Higgs field. To determine those relevant scales and therewith the Weinberg ratio
${g_1^2\over g_2^2}=\tan^2\th$
and the masses of the weak bosons,
 different strategies and model extensions
are proposed in the literature, 
e.g. the imbedding of $\U(1)$ and $\SU(2)$ as subgroups of
larger groups or substructures, etc.

In this paper the scales mentioned above are interpreted as symmetry 
normalizations which are 
determined by the representations of the relevant groups: $g_1^2,g_2^2$ as
normalizations of $\U(1)$ and $\SU(2)$,  the neutral weak boson mass
as normalization of the spin group in the Lorentz group 
$\O(3)\subnoteq\O(1,3)$
and the fine structure constant as normalization of the polarization group
$\O(2)\subnoteq\O(1,3)$. The ground state mass $M^2$ is related to the
normalization of the electroweak $\U(2)$. 

In addition to such a qualitative
interpretation of the electroweak scales 
a quantitive attempt is given to determine
with such a 
strategy the value of the Weinberg angle. 
This proposal relies on the experimental observation that the
electroweak group seems not to be a direct product,
but the product $\U(2)=\U(1_2)\o \SU(2)$ with the nontrivial 
 hypercharge-isospin correlation\cite{RAIF,S921}
 $\U(1_2)\cap\SU(2)=\{\pm\bl 1_2\}$. With an additional statistical
 argument a first approximation 
 of the normalization ratio 
 for the defining representation gives the value $\tan^2\th={1\over3}$.

\chapter{Masses and Coupling Constants}

In this chapter, the coupling constants and masses for the standard model
vector fields are given in a formulation adapted for an interpretation as group
normalization constants.

\section{Particle and Interaction Normalizations\\ in the Standard Model}
Obviously, coupling constants and masses play an important role in the
standard model interactions\cite{WEIN}. They allow
the experimental tests of the model by quantifying
 the transition from the interaction
oriented field language to the  ground state related 
particle language.

The standard model ground state properties 
are implemented by the complex $2$-dimensional
Higgs field $\phi^{\al=1,2}(x)$ with a mass $M$ as ground state value,
characterizing the minima of the potential $V(\phi)$
\begin{eqn}{l}
V(\phi)=\la_0(\phi^\star \phi-M^2)^2,~~\la_0>0\cr
\angle {\phi(x)}=\ul\phi={\scriptsize\pmatrix{
0\cr M\cr}},~~M\ne 0
\end{eqn}Therewith
a translation invariant  internal  reference system is fixed  
for the hyperisospin field symmetry
group $\U(2)=\U(1_2)\o\SU(2)$
up to a
 remaining abelian  particle symmetry with respect to 
 the electromagnetic $\U(1)_+\cong\U(1)$ subgroup 
\begin{eqn}{rll}
&\phi(x)\mape u(\al_0,\rvec\al)\phi(x),~~
&u(\al_0,\rvec\al)=e^{i{\al_0\bl 1_2+\rvec\al\rvec\tau\over\sqrt 2}}\in\U(2)\cr
&\ul\phi\mape 
u(\al_+)\ul\phi,~~&u(\al_+)=e^{i\al_+{\bl 1_2+\tau_3\over\sqrt2}}\in\U(1)_+\cr
\end{eqn}The electromagnetic $\U(1)_+$ is spanned with the projector
${\bl 1_2+\tau_3\over2}$ connecting hypercharge $\U(1_2)\cong\U(1)$ and
isospin $\SU(2)$.
With respect to the Killing form the  $\sqrt2$-normalization
is used in the $\U(2)$ Lie algebra basis 
$\{i{\bl 1_2\over \sqrt2},i{\rvec\tau\over\sqrt2}\}$ with the 
convenient 'double' trace
$\tr {\tau_a\over\sqrt2}{\tau_b\over\sqrt2}=\de_{ab}$.

With regard to the internal symmetries
the real $3$-dimensional Goldstone manifold $\U(2)/\U(1)_+$ 
characterizes the transition from fields to  particles. 
In contrast to the 
$\U(2)$-symmetry for fields, particles can have only a nontrivial 
electromagnetic $\U(1)$-symmetry. 
All particles come as $\SU(2)$-singlets - there is no mass degenerated nontrivial
particle isomultiplet.

The kinetic terms for the Higgs field 
in connection with the vector gauge fields 
$B^{j=0,1,2,3}(x)$
for hypercharge $\U(1_2)$ symmetry  and $\rvec W^j(x)$ for isospin $\SU(2)$ 
symmetry are given in the
 Lagrangian
\begin{eqn}{rl}
\bl L(\phi,B,W)
=&|\p_j-i{B_j+\rvec W_j\rvec\tau\over\sqrt2}\phi|^2- V(\phi)\cr
&+F^{jk}{\p_jB_k-\p_kB_j\over2}+g_1^2{F^{jk}F_{jk}\over4}\cr
&+\rvec F^{jk}{\p_j\rvec W_k-\p_k\rvec W_j\over2}+
g_2^2{\rvec F^{jk}\rvec F_{jk}\over4}
-{(\rvec F^{jk}\x\rvec W_k)\rvec W_j\over2}\cr
\end{eqn}The gauge fixing terms are omitted. For the vector fields a 1st 
order derivative formalism is used
with gauge fields and curvatures 
as canonical pairs, $(B,F)$ and $(\rvec W,\rvec F)$,
connected in the derivative terms. 
The curvatures arise by variation in the equations of motion
\begin{eqn}{l}
g_1^2F^{jk}=\p^kB^j-\p^jB^k,~~
g_2^2\rvec F^{jk}=\p^k\rvec W^j-\p^j\rvec W^k+\rvec W^k\x\rvec W^j\cr
\end{eqn}where the coupling constants show up 
as curvature normalizations. $g_1^2$ and $g_2^2$ are called internal
normalizations for hypercharge and isospin interaction resp.

A more familiar notation uses the coupling constants in the current gauge field
coupling with the rescaling 
\begin{eqn}{l}
\left.\begin{array}{ll}B= g_1\ul B,~~&F={1\over g_1}\ul F\cr
\rvec W= g_2\ul{\rvec  W},~~&\rvec F={1\over g_2}\ul{\rvec  F}
\end{array}\right\}
\then 
 J^kB_k+\rvec J^k\rvec W_k=
g_1J^k\ul B_k+g_2\rvec J^k\ul{\rvec W}_k
\end{eqn}which shows that the coupling constants are  
dynamically  relevant only for the case
of a nontrivial interaction. Only the squares $g^2$ occur,
therefore
$g=+\sqrt{g^2}$ can be defined as positive definite. Positive and negative
electromagnetic  charges arise by $\U(1)_+$-representations with positive or
negative winding numbers.

The hyperisospin transformation on the gauge sector is given by
\begin{eqn}{rll}
&B_j\mape B_j+\p_j\al_0,~~&F_{jk}\mape F_{jk}\cr
&\rvec W_j \mape O(\rvec \al)(\rvec W_j)+\p_j\rvec \al,~~
&\rvec F_{jk}\mape O(\rvec\al)(\rvec F_{jk})\cr
&&O(\rvec\al)\in\SO(3)\cong\U(2)/\U(1_2)\cr
\end{eqn}

The projection of the interaction fields to the ground state gives 
the vector particle  mass terms
\begin{eqn}{rl}
\bl L(\phi,B,W)=&|{\scriptsize\pmatrix{
B_j-W^3_j\cr
W_j^1+iW_j^2\cr}}|^2{M^2\over2}+\dots\cr
&+F^{jk}{\p_jB_k-\p_kB_j\over2}+g_1^2{F^{jk}F_{jk}\over4}
+\rvec F^{jk}{\p_j\rvec W_k-\p_k\rvec W_j\over2}+
g_2^2{\rvec F^{jk}\rvec F_{jk}\over4}+\dots
\end{eqn}diagonalizable with the Weinberg rotation
$O(\th)$ defining the massive vector field $Z^j(x)$ and the massless 
electromagnetic one $ A^j(x)$
\begin{eqn}{l}
O(\th)\!=\!{\scriptsize\pmatrix{
\cos\th&\tan\th\cr
-\sin\th&\cos\th\cr}}\in\SO(2):
\cases{ 
{\scriptsize\pmatrix{
\ul{F_A}\cr
\ul{F_ Z}\cr}} 
\!=\!{\scriptsize\pmatrix{
e F_A\cr
g_ZF_Z\cr}} 
\!=\!O(\th)
{\scriptsize\pmatrix{
g_1 F\cr
g_2 F_3\cr}}           
\!=\!O(\th)
{\scriptsize\pmatrix{
\ul{  F}\cr
\ul{  F_3}\cr}}                   \cr
{\scriptsize\pmatrix{
\ul A\cr
\ul Z\cr}}\!=\! 
{\scriptsize\pmatrix{
{1\over e} A\cr
{1\over g_Z} Z\cr}}
\!=\!O(\th) 
{\scriptsize\pmatrix{
{1\over g_1} B\cr
{1\over g_2}W_3\cr}} 
\!=\!O(\th) 
{\scriptsize\pmatrix{
\ul{B}\cr
\ul{W_3} \cr}}\cr} 
\end{eqn}The Weinberg angle $\th$ and the 
electromagnetic $\U(1)_+$-normalization (coupling constant) $ e^2$ are related
to the $\U(1_2)$ and $\SU(2)$ normalizations (coupling 
constants) $g_{1,2}^2$  by  two conditions:
The massive field $Z$ has
to occur in the combination of the $M^2$-proportional term 
in the Lagrangian and 
the massless field $A$ has to be 
normalized with respect to the remaining gauge invariance  
\begin{eqn}{rl}
Z_j\sim B_j-W_j^3,~~
&\then\cases{ 
\tan\th={g_1\over g_2}\cr
O(\th)={1\over\sqrt{g_1^2+g_2^2}}{\scriptsize\pmatrix{
g_2&g_1\cr
-g_1&g_2\cr}}\cr}\cr
 A_j\mape  A_j+\p_j\al_+&\then
{1\over  e^2}={1\over g_1^2}+{1\over g_2^2}\cr
\end{eqn}These two relations for $\th$ and $e^2$ define 
the electroweak rectangular triangle\cite{S921} with the $\U(1)_+$-hypotenusis
${1\over e}$ and  the $\U(1_2)$ and $\SU(2)$ sides ${1\over g_1}$ and ${1\over g_2}$
resp. The $Z$-coupling constant ${1\over g_Z}$ is determined as the
height of this triangle  
\begin{eqn}{l}
eg_Z=g_1g_2\then g_Z^2=g_1^2+g_2^2
\end{eqn}

The orthogonal Weinberg transformation $O(\th)$ for the rescaled fields $\ul B,\ul F$ etc.
is a special linear transformation 
$S(\th)$ for the unscaled fields $B,F$ etc.,
contragredient (inverse transposed, denoted by $-1T$) to each other
for the canonical pairs
\begin{eqn}{l}
\begin{array}{l}
S(\th)={\scriptsize\pmatrix{
1&1\cr
-\sin^2\th&\cos^2\th\cr}}\in\SL(\R^2)\cr
\d S(\th)=S(\th)^{-1 T}={\scriptsize\pmatrix{
\cos^2\th&\sin^2\th\cr
-1&1\cr}}\end{array}:
\cases{ 
{\scriptsize\pmatrix{
F_A\cr
F_ Z\cr}} 
=S(\th)
{\scriptsize\pmatrix{
F\cr
 F_3\cr}}\cr           
{\scriptsize\pmatrix{
A\cr
 Z\cr}} 
=\d S(\th) 
{\scriptsize\pmatrix{
 B\cr
W_3\cr}}\cr}
\end{eqn}The  transformation
$S(\th)$ shows the ground state induced
direction $\bl 1_2+\tau_3$ in the first line  as $(1,1)$, 
i.e. $F+F_3$, the contragredient
transformation $\d S(\th)$ involves the complementary 'broken'  direction
$\bl 1_2-\tau_3$ in the 2nd line $(-1,1)$, i.e. $-W_3+B$.

The normalizations (coupling constants) of the $\U(1_2)$ and $\SU(2)$ interaction
fields on the one side and ground state projected  fields on the
other side are related to each other as follows
\begin{eqn}{l}
S(\th){\scriptsize\pmatrix{
{1\over g_1^2}&0\cr 0&{1\over g_2^2}\cr}}S(\th)^T
=
{\scriptsize\pmatrix{
{1\over e^2}&0\cr 0&{1\over g_Z^2}\cr}}\hbox{ for }\tan^2\th={g_1^2\over g_2^2}
\end{eqn}

The Weinberg rotation from interaction fields to particle fields leads
to the kinetic terms for 
the vector fields
\begin{eqn}{rl}
\bl L(B,W)=& F_ A^{jk}{\p_j A_k-\p_k A_j\over2}
+ e^2{F_ A^{jk}F_{ A jk}\over4}\cr
&+F_Z^{jk}{\p_jZ_k-\p_kZ_j\over2}+
g_Z^2{F_Z^{jk}F_{Zjk}\over4}
  +{M^2(g_1^2+g_2^2)\over g_Z^2}{Z_jZ^j\over2}\cr
&+ {\SUM_{a=1,2}}(F_a^{jk}{\p_jW_{ak}-\p_kW_{aj}\over2}+
g_2^2{F^{jk}_a F_{ajk}\over4}+M^2{W_{aj}W^j_a\over2})
\end{eqn}

The particle masses $\mu$ arise from the products of
the coefficients for the curvature ${F^2\over 4}$ and the vector field 
${Z^2,W^2\over2}$ whereas the quotients of the coefficients 
define  constants 
$\ell^2$ and ${1\over\ell^2}$, called external normalizations related to the
particles 
\begin{eqn}{rl}
\bl L(B,W)=&F_ A^{jk}{\p_j A_k-\p_k A_j\over2}
+ e^2{F_ A^{jk}F_{ A jk}\over4}\cr
&+F_Z^{jk}{\p_jZ_k-\p_kZ_j\over2}+
\mu_Z(\ell^2_Z{F_Z^{jk}F_{Zjk}\over4}
  +{1\over\ell^2_Z}{Z_jZ^j\over2})\cr
&+ {\SUM_{a=1,2}}(F_a^{jk}{\p_jW_{ak}-\p_kW_{aj}\over2}+
\mu_W(\ell^2_W{F^{jk}_a F_{ajk}\over4}+{1\over\ell^2_W}{W_{aj}W^j_a\over2}))\cr
\hbox{with }&\left\{
\begin{array}{ll}
\mu^2_Z=(g_1^2+g_2^2)M^2=g_Z^2M^2,~~
&\ell^4_Z={g_1^2+g_2^2\over M^2}={g_Z^2\over M^2}\cr
\mu^2_W=g_2^2M^2,~~&\ell^4_W={g_2^2\over M^2}\end{array}\right.\cr
\end{eqn}Planck's constant 
$\hbar$ and the maximal action 
velocity $c$ are used as natural units.

The currents $J^k(x)$ for hypercharge $\U(1_2)$ and $\rvec J^k(x)$ 
for isospin $\SU(2)$ have to be transformed as the curvatures
\begin{eqn}{rll}
{\scriptsize\pmatrix{
J_A\cr
J_Z\cr}} 
&=S(\th)
{\scriptsize\pmatrix{
 J\cr
 J_3\cr}}&={\scriptsize\pmatrix{
J+J_3\cr
-\sin^2\th J+\cos^2\th J_3\cr                  
}}\cr
\end{eqn}to obtain the interaction in the asymptotic particle basis 
\begin{eqn}{rl}
-\bl L(J,B,W)&=J^kB_k+\rvec J^k\rvec W_k=
J_ A^k  A_k+ J_Z^k Z_k+{\SUM_{a=1,2}}J_a^kW_{ak}\cr
&=(J^k+J^k_3)A_k+(-\sin^2\th J^k+\cos^2\th J_3^k)Z_k
+{\SUM_{a=1,2}}J_a^kW_{ak}\cr
\end{eqn}

The standard model cannot determine the  dimensionless values
${ e^2\over4\pi}\sim{1\over137}$, $\tan^2\th\sim0.3$ and 
the ratio of $M\sim 123 {\ro{GeV}\over c^2}$ to any reference mass.
 Those values  are taken from the experiments with the energy momentum 
 dependent Weinberg angle at the $Z$-mass. 
 Using dual 
 rectangular triangles 
 \begin{eqn}{l}
 \left.\begin{array}{rl}
\cl T=  (a_1,a_2;c|h):~~&a_1^2+a_2^2=c^2\cr
 \d{\cl T}= ({1\over a_2},{1\over a_1};{1\over h}|{1\over c}):~~
 &{1\over a_2^2}+{1\over a_1^2}={1\over h^2}\end{array}\right\}
 ,~~a_1a_2=ch,~~\tan\th={a_1\over a_2}
 \end{eqn}one has for the coupling constants $\cl G$,  the
 masses $\cl M$ and the normalization constants $\cl L^2$
 \begin{eqn}{rl}
\hbox{couplings:}&
 \cl G=(g_1,g_2;g_Z|e)
 \sim({1\over 2.9},{1\over 1.6};{1\over 1.4}|{1\over 3.3})\cr
\hbox{masses:}&\cl M=(\mu_1,\mu_W;\mu_Z|\mu_e)
\sim(43.4,80.2;91.2|38.2){\ro{GeV}\over c^2}\cr
\hbox{normalizations:}&\cl L^2=(\ell_1^2,\ell_2^2;\ell_Z^2|\ell_ e^2) \cr
&\cl M=M\cl G,~\cl L^2={1\over M}\cl G\cr
\end{eqn}$\mu_e$ and $\mu_1$
are no particle masses
in contrast to the $W$- and $Z$-masses $\mu_W$ and $\mu_Z$ resp.

The {\it internal interaction
normalizations} are $\cl G^2$ and ${1\over \cl G^2}$,
the {\it external particle normalizations} $\cl L^2$ and ${1\over\cl L^2}$,
related to each other by the ground state mass value $M$.

The  Feynman
propagators for the particle vector fields
are characterized by the mass as momentum space pole  and the 
residue as the coupling constant involved
\begin{eqn}{rll}
\angle{\cl T Z^kZ^j}(x)&=
-{i\over\pi}\int {d^4 q e^{iqx}\over(2\pi)^3}(\eta^{kj}-{q^kq^j\over \mu_Z^2})
{g_Z^2\over q^2+io-\mu_Z^2},&g_Z^2={\mu_Z^2\over M^2}=\mu_Z\ell_Z^2\cr
\angle{\cl T A^kA^j}(x)&=
-{i\over\pi}\int {d^4 qe^{iqx}\over(2\pi)^3}\eta^{kj}
{ e^2\over q^2+io}+\hbox{gauge dep.},& e^2={\mu_ e^2\over M^2}=\mu_e\ell_e^2\cr
\angle{\cl T W_a^kW_b^j}(x)&=
-\de_{ab}{i\over\pi}\int {d^4 qe^{iqx}\over(2\pi)^3}(\eta^{kj}-{q^kq^j\over \mu_W^2})
{g_2^2\over q^2+io-\mu_W^2},&g_2^2={\mu_W^2\over M^2}=\mu_W\ell_W^2\cr
\end{eqn}

\section{The Quantum Mechanical Oscillators}

The quantum mechanical oscillators can be used as a simple model 
to illustrate the interplay 
of masses and coupling constants.

The isotropic Bose oscillator in $n$ dimensions
has as Hamiltonian with mass $m$ and spring constant $k$
\begin{eqn}{l}
H_n={\rvec p^2\over 2m}+k{\rvec q^2\over 2},~~(\rvec p,\rvec q)=(p_a,q_a)_{a=1}^n
\end{eqn}to be compared with the field theoretical expression
$g^2{F^2\over4}+M^2{W^2\over2}$ e.g. for the massive $W_{1,2}$-fields.
The quantum mechanical analogue of the coupling constants $g^2$ for
relativistic fields and the ground state mass $M^2$ is the inverse inert
mass
${1\over m}$ for mass points and the spring constant $k$ resp.
The characteristic frequency $\mu$ and the intrinsic  length $\ell$ 
of the oscillator  are
given with the product and the quotient of inverse mass and spring constant
\begin{eqn}{l}
H_n=\mu{\ell^2\rvec p^2+{1\over\ell^2}\rvec q^2\over2},~~
\mu^2={k\over m},~~\ell^4={1\over mk}
\end{eqn}$\hbar$ is used as natural unit.
The mass of a  particle, e.g. the $Z$ or $W$ mass  $\mu_{Z,W}$, has its
analogue in the mass point frequency $\mu$, not in the inert mass $m$.

The time dependent quantization of the harmonic oscillator
with the shorthand notation 
$\com ab(t-s)=\com{a(s)}{b(t)}$ is a $\U(1)\cong\SO(2)$ 
representation
\begin{eqn}{l}
D_{ab}(t|\mu)={\scriptsize\pmatrix{
\com{i p_a}{q_b}&
\com{q_a}{q_b}\cr
\com{ p_a}{p_b}&
\com{q_a}{-ip_b}\cr}}(t)=
\de_{ab}D_1(t|\mu)\cr
~~\cr
D_1(t|\mu)={\scriptsize\pmatrix{
\cos t\mu& i\ell^2\sin t\mu\cr
{i\over\ell^2}\sin t\mu&\cos t\mu\cr}}=e^{it\mu N(\ell^2)},~~N(\ell^2)
={\scriptsize\pmatrix{
0& \ell^2\cr
{1\over\ell^2}&0\cr}}  
\end{eqn}

The harmonic oscillator  mass arises in the residue ${1\over m}$ of the
integral representation for the time dependent quantization. One has 
with the principial 
value $(\mu^2)_P$ integration contour
in the energy plane 
\begin{eqn}{l}
\com {q_a}{q_b}(t)=
\de_{ab}i\ell^2\sin t\mu=-\de_{ab}{i\ep(t)\over\pi}
\int dE e^{it E}{{1\over m}\over
E^2-(\mu^2)_P},~~~{1\over m}={\mu^2\over k}=\mu\ell^2
\end{eqn}

The characteristic length $\ell$ can be absorbed 
in the 'free' oscillator by the rescaling
\begin{eqn}{l}
\rvec q=\ell\ul{\rvec q},~~\rvec p={1\over\ell}\ul{\rvec p}
\end{eqn}i.e. it has no physical relevance.
If, however, a nontrivial 
rotation $\O(n)$-invariant $\rvec q^2$-dependent potential $V$ is added,
e.g.  $(\rvec q^2)^2$  for an  anharmonic oscillator or
$-k{\rvec q^2\over2}-{1\over\sqrt{\rvec q^2}} $ for the Kepler dynamics,
$\ell^2$ is relevant for the coupling constant
\begin{eqn}{rl}
H_n+V&= 
\mu{\ell^2\rvec p^2+{1\over\ell^2}\rvec q^2\over2}+V(\rvec q^2)=
\mu{\ul{\rvec p}^2+\ul{\rvec q}^2\over2}+V(\ell^2\ul{\rvec q}^2)\cr
\end{eqn}

The oscillator length $\ell$  normalizes the 'real' and 'imaginary'
combination of the creation operator $\ro u$ and the annihilation operator
$\ro u^\star$ 
\begin{eqn}{l} 
\rvec q=\ell{\rvec{\ro u}+\rvec {\ro u}^\star\over\sqrt2},~~
\rvec p={i\over\ell}{\rvec {\ro u}-\rvec {\ro u}^\star\over\sqrt2}\cr
\end{eqn}In the complex creation-annihilation formulation the
characteristic length does not show up in the $\U(n)$-symmetric Hamiltonian
\begin{eqn}{l}
H_n=\mu{\acom{\rvec \ro u}{\rvec {\ro u}^\star}\over2},~~
\com{\ro u^\star_ a}{\ro u_b}(t)
=\de_{ab}e^{it\mu}
\end{eqn}

The quantum mechanical meaning of the 
particle normalization $\ell^2$
is given with the Fock ground state of the harmonic oscillator.
For the Fock space with the ground state values
\begin{eqn}{l}
n=1:~~\angle{(\ro u^\star \ro u)^N}=1,~~N=0,1,\dots\cr
n\ge1:~~\angle{
(\ro u^\star_{a_1}\ro u_{b_1})
\cdots( \ro u^\star_{a_N}\ro u_{b_N})}=\de_{a_1b_1}\cdots\de_{a_Nb_N}
\end{eqn}the time dependent Fock values for position and momenta
with the shorthand  notation 
$\angle{\acom ab}(t-s)=\angle{\acom{a(s)}{b(t)}}$
are the product of the particle normalization matrix $N(\ell^2)$ with a time
representation
\begin{eqn}{l}
{\scriptsize\pmatrix{
\angle{\acom{i p_a}{q_b}}&\angle {\acom{q_a}{q_b}}\cr
\angle{\acom{ p_a}{p_b}}&\angle {\acom{q_a}{-ip_b}}\cr}}(t)=
\de_{ab}{\scriptsize\pmatrix{
i\sin t\mu& \ell^2\cos t\mu\cr
{1\over\ell^2}\cos t\mu &i\sin t\mu\cr}}=\de_{ab}N(\ell^2) D_1(t|\mu)\cr
{\scriptsize\pmatrix{
\angle{\acom{i p_a}{q_b}}&\angle {\acom{q_a}{q_b}}\cr
\angle{\acom{ p_a}{p_b}}&\angle {\acom{q_a}{-ip_b}}\cr}}(0)=
\de_{ab}N(\ell^2)=\de_{ab}{\scriptsize\pmatrix{
0& \ell^2\cr
{1\over\ell^2} &0\cr}}
\end{eqn}$N(\ell^2)$ contains the dual normalizations of positions
and momenta, i.e. $\ell^2$ and  ${1\over\ell^2}$ resp.

The sum of 
time ordered  quantization and Fock values is the
quantum mechanical analogue of the Feynman propagator, given
for the position with
the oscillator frequency as energy pole and the inverse mass as residue
\begin{eqn}{rl}
\angle{\cl T qq}(t)&=\angle{\acom qq}(t)-\ep(t)\com qq(t)
=\ell^2(\cos t\mu-i\ep(t)\sin t\mu)\cr
&=\ell^2 e^{-i|t|\mu}=-{i\over\pi}\int dE e^{itE}{{1\over m}\over E^2+io-\mu^2}
,~~{1\over m}={\mu^2\over k}=\mu\ell^2\end{eqn}

\section{Fock Normalization for Particles}

The particle expansion of the vector 
fields in the standard model\cite{SBH}
\begin{eqn}{rl}
Z^j(x)&={\SUM_{n=1,2,3}}\int{d^3q\over(2\pi)^3}
\sqrt{{\mu_Z\over q_0}}\La({q\over\mu_Z})^j_n
~\ell_Z{e^{ixq}\ro u^n(\rvec q)+e^{ -ixq}\ro u^{\star n}(\rvec q)\over\sqrt2}\cr
W_a^j(x)&={\SUM_{n=1,2,3}}\int{d^3q\over(2\pi)^3}
\sqrt{{\mu_W\over q_0}}\La({q\over\mu_W})^j_n
~\ell_W{e^{ixq}\ro u_a^n(\rvec q)+e^{ -ixq}\ro u_a^{\star n}(\rvec q)\over\sqrt2}
,~~a=1,2\cr
A^j(x)&={\SUM_{n=0,1,2,3}}\int{d^3q\over(2\pi)^3}
\sqrt{{ \mu_e\over q_0}}H({\rvec q\over q_0})^j_n
~\ell_e {\scriptsize\pmatrix{
\dots\dots\cr
~\cr
{e^{ixq}\ro U^1(\rvec q)+e^{ -ixq}\ro U^{\star 1}(\rvec q)\over\sqrt2}\cr
{e^{ixq}\ro U^2(\rvec q)+e^{ -ixq}\ro U^{\star 2}(\rvec q)\over\sqrt2}\cr
~\cr
\dots\dots\cr}}\cr
\end{eqn}uses the energy $q_0=\sqrt{\mu^2+\rvec q^2}$ for
$\mu=\mu_Z,\mu_W,0$ and the boost $\La({q\over\mu})$ 
with positive mass $\mu>0$ for $\SO^+(1,3)/\SO(3)$ 
to transform to a rest system and 
$H({\rvec q\over q_0})$,  $q_0=|\rvec q|$, i.e. with vanishing mass $\mu=0$, 
for $\SO^+(1,3)/\SO(2)$ to transform to a rest system with a polarization
direction.  
The two nonparticle components $(\dots)$ 
(Coulomb and gauge degree of freedom for $n=0,3$) 
for the massless electromagnetic
field are not given explicitly.  

The quantization and Fock values for the creation and annihilation operators
\begin{eqn}{l}
\com{\ro u^\star(\rvec p)}{\ro u(\rvec q)}=\de(\rvec q-\rvec p)=  
\angle {\acom{\ro u^\star(\rvec p)}{\ro u(\rvec q)}}
\end{eqn}exhibit  the external normalizations
$\ell_Z^2,\ell_W^2$ for the massive particles involved
\begin{eqn}{rl}
\angle{\acom{Z^k}{Z^j}}(x)&=
{\SUM_{n,m=1,2,3}}
\int{d^3qe^{-i\rvec x\rvec q}\over(2\pi)^3}    
{\mu_Z\over q_0}
\La({q\over\mu_Z})^k_m
~\ell_Z^2\de^{mn}\cos x_0q_0~
\La({q\over\mu_Z})^j_n\cr
\angle{\acom{W_a^k}{W_b^j}}(x)&=
{\SUM_{n,m=1,2,3}}
\int{d^3qe^{-i\rvec x\rvec q}\over(2\pi)^3}
~{\mu_W\over q_0}
\La({q\over\mu_W})^k_m
~\ell_W^2\de^{mn}\de_{ab}\cos x_0q_0~ 
\La({q\over\mu_W})^j_n\cr
\end{eqn}The Fock values for the two massless particle degrees of freedom
(photons) $m,n=1,2$ in the  electromagnetic  field
\begin{eqn}{l} 
\angle{\acom{A^k}{A^j}}(x)=
{\SUM_{n,m=0,1,2,3}}
\int{d^3qe^{-i\rvec x\rvec q}\over(2\pi)^3}    
{\mu_e\over q_0}H({\rvec q\over q_0})^k_m
~\ell_e^2{\scriptsize\pmatrix{ 
\cdots & 0&\cdots\cr
0&\cos x_0q_0&0\cr
\cdots&0&\dots\cr}}^{mn}
H({\rvec q\over q_0})^j_n\cr
\end{eqn}have as external particle normalization $\ell_ e^2={ e^2\over \mu_e}$.

\chapter{Normalization of Symmetries}
In this chapter, the mathematical structure for  symmetry normalizations
(coupling constants and particle masses) is given together with its application to the standard
model vector fields.

\section{Dual Normalizations}

A nondegenerate  quadratic form $\ze$, i.e. a  symmetric  real bilinear  or
complex sesquilinear form of a finite dimensional vector space $V$
with scalars $\K=\R$ and $\K=\C$ resp.
\begin{eqn}{l}
\ze(~,~):V\x V\map \K,~~\cases{
\ze(v,w)=\ze(w,v)\hbox{, bilinear}\cr 
\ze(v,w)=\ol{\ze(w,v)}\hbox{, sesquilinear}\cr }
\end{eqn}induces
an (anti)linear isomorphism to the dual space $V^T$, 
containing  the linear $V$-forms
\begin{eqn}{l}
\ze:V\map V^T, v\mape \ze(v)=\ze(v,~),~~\cases{
\ze(v+w)=\ze(v)+\ze(w)\cr
\ze(\al v)=\al\ze(v),~\al\in\R\hbox{, bilinear}\cr
\ze(\al v)=\ol\al\ze(v),~\al\in\C\hbox{, sesquilinear}\cr}
\end{eqn}and therewith the inverse quadratic form $\d\ze$ for
the linear forms
\begin{eqn}{l}
\d\ze(~,~):V^T\x
V^T\map\K,~~\d\ze(\om,\th)=\ze(\ze^{-1}(\om),\ze^{-1}(\th))
\end{eqn}

With the isomorphism $\ze$, the quadratic form 
can be expressed by the dual product
of vectors $V$ and linear forms $V^T$
\begin{eqn}{rll}
\hbox{dual product}:&V^T\x V\map \K,~~&(\om,v)\mape\om(v)=\dprod\om v\cr
\ze(~,~):&V\x V\map\K,~~&\ze(v,w)=\dprod{\ze(v)}w\cr
\d\ze(~,~):&V^T\x V^T\map\K,~~&\d\ze(\om,\th)=\dprod\om{\ze^{-1}(\th)}\cr
\end{eqn}

Nondegenerate quadratic forms are 
characterized by a signature $(n_+,n_-)$,
i.e. by their real orthogonal 
invariance groups $\O(n_+,n_-)$ for real vector spaces 
$V\cong\R^n$, $n=n_++n_-$, or unitary ones $\U(n_+,n_-)$
for complex vector spaces $V\cong\C^n$.
If $\{e^a\}_{a=1}^n$ and
$\{\d e_a\}_{a=1}^n$ are dual bases of $V$ and $V^T$, i.e.
with the dual product $\dprod{\d e_a}{e^b}=\de_a^b$, their squares
build the matrices
 $\ze(e^a,e^b)=\ze^{ab}$ and $\d\ze(\d e_a,\d e_b)=\ze_{ab}$.

A quadratic form of a complex vector space defines a 
$\U(n_+,n_-)$-conjugation
\begin{eqn}{l}
\bl U(n_+,n_-):~~\ze(v)=v^*=\ze(v,~),~~\ze^{-1}(\om)=\om^*=\d\ze(\om,~)
\end{eqn}

 Only  real, not complex quadratic forms are expressable by 
power two tensors, written in, but independent of bases
\begin{eqn}{l}
\bl O(n_+,n_-):~~\left\{\begin{array}{rl}
\d\ze&=\ze_{ab} e^a\ox e^b\in V\ox V\cr
\ze&=\ze^{ab} \d e_a\ox \d e_b\in V^T\ox V^T\end{array}
\right.
\end{eqn}

Orthogonal  bases of vector spaces with definite bilinear forms, i.e. 
of scalar products with $n_-=0$, define
dual normalizations $\ell^2$ and ${1\over\ell^2}$ for the 
compact symmetries acting on  the vector spaces $V$ and $V^T$
\begin{eqn}{rll}
&\hbox{for }\O(n),\U(n):~~
&\left\{\begin{array}{rl}
\ze(e^a,e^b)&= \ell^2\de^{ab} \cr  
\d\ze(\d e_a,\d e_b)&={1\over\ell^2}\de_{ab}\cr
\end{array}\right.,~~\ell^2>0\cr
&\hbox{for }\O(n):~~&\left\{\begin{array}{rl}
\d\ze&={1\over\ell^2}\de_{ab} e^a\ox e^b\in V\ox V\cr
\ze&=\ell^2 \de^{ab} \d e_a\ox \d e_b\in V^T\ox V^T\end{array}
\right.
\end{eqn}

\section{Dual Normalizations for Oscillators}

The harmonic Bose oscillator (section 1.2) has
an $\O(n)$-invariant scalar product for the dual spaces 
 of positions $q\in\S\cong\R^n$  and momenta $p\in\S^T\cong\R^n$
\begin{eqn}{rll}
&\si(~,~):\S\x\S\map \R,~~&\cases{
\si(q,q')=\si(q',q)\cr
\si(q,q)>0\iff q\ne0\cr}\cr
&\d\si(~,~):\S^T\x \S^T\map \R,~~&\cases{\d\si(p,p')=\d\si(p',p)\cr
\d\si(p,p)>0\iff p\ne0\cr}\cr
\end{eqn}Using dual and orthonormal bases $\{q_a,p_a\}_{a=1}^n$
one has the invariant tensors with the intrinsic oscillator lenght $\ell$
\begin{eqn}{l}
\dprod{p_a}{q_b}=\de_{ab},~~\left\{\begin{array}{rll}
&\si(q_a,q_b)=\ell^2\de_{ab}&\then
 \si=\ell^2 p_a\ox p_a\cr
& \d\si(p_a,p_b)={1\over\ell^2}\de_{ab}&\then
\d \si={1\over\ell^2} q_a\ox q_a\cr
 \end{array}\right.
\end{eqn}The sum of the dual scalar products, multiplied  
 with a constant $\mu$ leads to the quantum 
mechanical   Hamiltonian
\begin{eqn} {rl}
\dprod{p_a}{q_b}=\de_{ab}&\then
\com{ip_a}{q_b}=\de_{ab}\cr
\mu{\d\si+\si\over2}=
\mu{{1\over \ell^2}p_a\ox p_a+\ell^2 q_a\ox q_a\over 2}&\then H_n 
=\mu{{1\over \ell^2}p_a p_a+\ell^2 q_aq_a\over 2}\cr
\end{eqn}$\ell^2$ and ${1\over\ell^2}$ are the dual 
particle normalizations of $\O(n)$
acting on positions and momenta.

\section{Interaction Normalizations\\for the Standard  Gauge Fields}
 
With respect to the
internal degrees of freedom 
the gauge field sector in the standard model is built with dual 
real 4-dimensional vector spaces for gauge vector fields and 
curvatures
which are the direct sums of 1- and 3-dimensional subspaces
\begin{eqn}{rl}
\hbox{curvatures: } G= G_1\pl  G_3\cong\R\pl\R^3&\hbox{with basis }\{F,F_a\}\cr 
\hbox{gauge vectors: } G^T= G_1^T\pl  G_3^T\cong\R\pl\R^3&\hbox{with basis }\{B,W_a\}\cr 
\hbox{dual bases: }\dprod B F=1,&~~\dprod {W_a}{F_b}=\de_{ab}
\end{eqn}The hyperisospin group $\U(2)=\U(1_2)\o\SU(2)$ acts nontrivially 
only on the
isospin subspaces $ G _3, G _3^T$ 
as $\U(2)/\U(1_2)\cong\SO(3)$. 

The squares of the coupling constants reflect a $\U(2)$-invariant
scalar product with dual normalizations 
$g^2$ and ${1\over g^2}$ of the internal groups 
acting on gauge vectors and curvatures
\begin{eqn}{rll}
 g(~,~): G \x  G \map \R,& g(F,F)={1\over g_1^2}
,&g(F_a,F_b)=\de_{ab}{1\over g_2^2}\cr
\d g(~,~): G ^T\x  G ^T\map \R,&\d g(B,B)=g_1^2,&\d g(W_a,W_b)=\de_{ab}g_2^2\cr
\d g=g_1^2 F^2+g_2^2\rvec F^2,&
 g={1\over g_1^2}B^2+{1\over g_2^2}\rvec W^2
 \end{eqn}The Lagrangian 
 uses the gauge invariant metrical tensor $\d g$ for the curvatures.
 
The distinction of the electromagnetic group $\U(1)_+$-direction $\bl 1_2+\tau_3$ 
in the  ground state defines
the basis transformation  $S(\th)\in\SL(\R^2)$ in the 2-dimensional
curvature subspace spanned by $\{F,F_3\}$ 
and to the contragredient transformation
in the vector field space spanned by $\{B,W_3\}$ (section 1.1)
\begin{eqn}{lll}
\Th: G \map G ,&
{\scriptsize\pmatrix{
F_A\cr F_ Z\cr F_{1,2}\cr}}=\Th
{\scriptsize\pmatrix{F\cr F_3\cr F_{1,2}\cr}},&
\Th={\scriptsize\pmatrix{
1&1&0\cr
-\sin^2\th&\cos^2\th&0\cr 0&0&\bl 1_2\cr}}\in\SL(\R^4)\cr
\d \Th: G ^T\map G ^T,&{\scriptsize\pmatrix{
A\cr  Z\cr W_{1,2}\cr}}=
\d \Th{\scriptsize\pmatrix{B\cr W_3\cr W_{1,2}\cr}},&
\d \Th={\scriptsize\pmatrix{
\cos^2\th&\sin^2\th&0\cr
-1&1&0\cr 0&0&\bl 1_2\cr}}\in\SL(\R^4)
\end{eqn}The internal normalizations read in the new basis
\begin{eqn}{rl}
\Th{\scriptsize\pmatrix{
{1\over g_1^2}&0\cr
0&{1\over g_2^2}\bl 1_3\cr}}\Th^T&=
{\scriptsize\pmatrix{
{1\over e^2}&0&0\cr
0&{1\over g_Z^2}&0\cr
0&0&{1\over g_2^2}\bl 1_2\cr}}\cr
\d g=g_1^2 F^2+g_2^2\rvec F^2&=e^2 F_A^2+g_Z^2F_Z^2+g_2^2{\SUM_{a=1,2}}F_a^2
\end{eqn}

\section{Higgs Mass as Unit\\for  the Goldstone Manifold}

The algebraic framework for the 
ground state properties implementing complex Higgs vectors
$\phi$ (section 1.1)
uses  a $\U(2)$-invariant 
scalar product of the complex 2-dimensional Higgs vector space $\H\cong\C^2$ 
\begin{eqn}{l}
\sprod{~}{~}:\H\x \H\map \C,~~\cases{
\sprod{\phi}{\phi'}=
\ol{\sprod{\phi'}{\phi} }\cr
\sprod{\phi}{\phi}>0\iff\phi\ne 0\cr
\phi^\star=\sprod\phi{~}\in\H^T\cr}
\end{eqn}

In the  real $4$-dimensional manifold $\GL(\C^2)_\R/\U(2)$  of the
 classes of 
 $\U(2)$-equivalent bases with $\GL(\C^2)_\R$ taken as real $8$-dimensional
Lie group, the orthogonal bases 
$\{\phi^\al\}_{\al=1,2}$  
with the diagonal matrix
\begin{eqn}{l}
\sprod{\phi^\al}{\phi^\be}=M^2\de^{\al\be},~~
\sprod{~}{~}\cong M^2{\scriptsize\pmatrix{1&0\cr 0&1\cr}}
\end{eqn}define the normalization $M^2$  of the 
hyperisospin group $\U(2)$ acting on the Higgs space.

Any nontrivial vector $\ul\phi\in\H$ fixes an orthogonal basis up to 
the electromagnetic $\U(1)_+$-direction. Therefore $M$ can also be called
the unit of the Goldstone manifold $\U(2)/\U(1)_+$ which relates fields and
particles with repect to the internal degrees of freedom.

\section{Induced Normalizations \\for Lie Algebras}
 
If a real Lie algebra $L$ acts 
on  a real or complex vector space $V$   by linear endomorphisms
\begin{eqn}{l}
L\x V\map V,~~ l\m v=\cl D(l)(v)\cr
\end{eqn}each vector $v\in V$ induces via a
quadratic form $\ze$ of $V$  a 
symmetric bilinear form on the Lie algebra, 
definite for a scalar product
\begin{eqn}{l}
\ze_v(~,~):L\x L\map \R,~\cases{
\ze_v(l,l')=\ze(l\m v,l'\m v) 
=\ze_v(l',l)\cr
\hbox{if }\ze(v,v)> 0\hbox{ for }v\ne0\then \ze_v(l,l)\ge 0\cr}
\end{eqn}The induced  bilinear Lie algebra form is trivial for the
Lie subalgebra which acts trivially on the vector $v$
\begin{eqn}{l}
\ze_v(l,L)=\{0\}~~\hbox{ for }~~l\in\FIX_vL
=\{l\in L\mid l\m v=0\}\cr
\end{eqn}$\ze_v$ is nondegenerate for the quotient vector space 
$L/\FIX_v L$ and a scalar product thereon  for a scalar product $\ze$.

In the standard  model (section 1.1)
the vector fields obtain their masses in the
mechanism just described:
The Lie algebras for
hypercharge and for isospin  
$\log\U(2)=\log\U(1_2)\pl\log\SU(2)$ act on the
Higgs vectors $\H\cong\C^2$
\begin{eqn}{rll}
&l^0\m\phi^\al={i\over\sqrt2}\phi^\al,~~&l^0\in\log\U(1)\cong\R\cr
&l^a\m\phi^\al={i\over\sqrt 2}(\tau^a)^\al_\be\phi^\be,~~
&l^a\in\log\SU(2)\cong\R^3\cr
\end{eqn}The electromagnetic Lie algebra 
$\log\U(1)_+$ is the Lie algebra acting trivially on each 
nontrivial Higgs vector
\begin{eqn}{l}
\H\ni\ul\phi\cong{\scriptsize\pmatrix{0\cr M\cr}}\ne 0
,~~\sprod{\ul\phi}{\ul\phi}=M^2\cr
\FIX_{\ul\phi}\log\U(2)=\log\U(1)_+
\cong i\R{\bl 1_2+\tau_3\over\sqrt2}\cr
\end{eqn}The  scalar product
induced on the Lie algebra $\log\U(2)$  is nontrivial on the
vector space $\log \U(2)/\U(1)\cong\R^3$ for the massive $Z$- and 
$W_{1,2}$-particle fields
\begin{eqn}{rll}
\sprod{l^0-l^3}{l^0-l^3}_{\ul\phi}&=
\sprod{i{\bl 1_2-\tau^3\over\sqrt2}}{i{\bl 1_2-\tau^3\over\sqrt2}}_{\ul\phi}
&={M^2\over2}\cr
\sprod{l^a}{l^b}_{\ul\phi}&=
 \sprod{i{\tau^a\over\sqrt2}}{i{\tau^b\over\sqrt2}}
 _{\ul\phi}&=\de_{ab}{M^2\over2}\hbox{ for } 
a,b=1,2\cr
\end{eqn}

\section{Particle Normalization\\for the Standard Gauge Fields}

The
normalization of the Lorentz group $\SO^+(1,3)$-action shows up in the
particle normalizations $\ell^2$ (section 1.1). 
The distinction of the definite subgroups of the Lorentz group
goes parallel with the Wigner classification\cite{WIG,SBH,S961}
 of the representations of
the Poincar\'e group: Via the massive particles  the definite spin group 
$\SO(3)\cong\SU(2)/\I_2$ is normalized, via
the massless ones the definite cirularity
(polarization) group $\SO(2)\cong\U(1)$ inside the Lorentz group
$\SO^+(1,3)\cong\SL(\C^2)_\R/\I_2$.

Similar to the internal symmetries with the 
Goldstone manifold $\U(2)/\U(1)_+$ 
characterizing the field particle transition, there arise 
for the external symmetries the Sylvester and
Witt
manifolds $\SO^+(1,3)/\SO(3)$ and $\SO^+(1,3)/\SO(2)$ resp.
relevant for the harmonic particle analysis of the relativistic fields.
In contrast to the fields, particles have no full Lorentz symmetry, massive particles keep the
spin $\SO(3)$ symmetry, massless ones only the circularity 
(polarization) $\SO(2)$ symmetry.

To be more explicit:
The  Lagrangian for a massive  vector field
\begin{eqn}{l}
\bl L(Z,F_Z)={1\over2}F_Z^{jk}\ep^{lm}_{jk}\p_l Z_m
+\mu_Z(\ell_Z^2{F_Z^{jk}F_{Zjk}\over 4}+{Z^jZ_j\over2\ell_Z^2})
\end{eqn}with the Clebsch-Gordan coefficients 
$\ep^{lm}_{jk}=\de^l_m\de^m_k-\de^l_k\de^m_j$  and
dual normalization constant $\ell_Z^2$ and ${1\over\ell_Z^2}$
gives rise to the commutators of the canonical pair $(Z,F_Z)$
\begin{eqn}{rl}
&{\scriptsize \pmatrix{
\com{iF_Z^{kl}}{Z^j}&\com {Z^k}{Z^j}\cr
\com{F_Z^{kl}}{F_Z^{jm}}&\com {Z^k}{-iF_Z^{jm}}\cr}}(x)\cr
&~~\cr
&={\scriptsize \pmatrix{
-i\ep_{ut}^{lk}\de^j_s{\p^u\over \mu_Z} &\ell_Z^2\de_t^k\de_s^j\cr
-{1\over\ell_Z^2}\ep_{ut}^{lk}\ep_{rs}^{mj}{\p^r\p^u\over \mu_Z^2}
&-i\de^k_t\ep_{rs}^{mj}{\p^r\over \mu_Z}\cr}}
\int {d^4 qe^{ixq}\over(2\pi)^3}\ep(q^0)\mu_Z\de(q^2-\mu_Z^2)
(-\eta^{kj}+{q^kq^j\over\mu_Z^2})\cr
\end{eqn}
The $Z$-commutators
\begin{eqn}{l}
\com {Z^k}{Z^j}(x)
= \int {d^3 q e^{-i\rvec x\rvec q}\over(2\pi)^3}{\mu_Z\over q_0}~
\La({q\over\mu_Z})_m^k\brack{ZZ}^{mn}(x_0)
\La({q\over \mu_Z})^j_n\cr
\end{eqn}involve the $\SO(3)$ normalization 
$\ell_Z^2$ 
\begin{eqn}{l}
\brack{ZZ}(x_0)=
\ell_Z^2~{\scriptsize \pmatrix{0&0\cr 0&\bl 1_3\cr}}i\sin x_0q_0
,~~\ell_Z^2={g_Z^2\over\mu_Z}={\mu_Z\over M^2} \cr
\end{eqn}and the boost $\La({q\over\mu_Z})$ for the 
transformation to a Sylvester basis
\begin{eqn}{l}
\La({q\over\mu_Z})\cong{1\over\mu_Z}{\scriptsize\pmatrix{
q_0&\rvec q\cr \rvec q& \de_{ab}\mu+{q_aq_b\over q_0+\mu_Z}\cr}},~~q^2=\mu_Z^2
\end{eqn}The boosts represent $\SO(3)$-classes of the
Sylvester manifold $\SO^+(1,3)/\SO(3)$.
The Fock values of the anticommutators use the analogue
structures with 
\begin{eqn}{l}
\angle{\acom {Z^k}{Z^j}}(x)
= \int {d^3 qe^{-i\rvec x\rvec q}\over(2\pi)^3}{m\over q_0}~
\La({q\over\mu_Z})_m^k
\angle{\brace{ZZ}^{mn}}(x_0)
\La({q\over \mu_Z})_n^j\cr
\angle{\brace{ZZ}}(x_0)=
\ell_Z^2~{\scriptsize \pmatrix{0&0\cr 0&\bl 1_3\cr}}\cos x_0q_0\cr
\end{eqn}For  massive vector particles a rest system with a time direction
has to be fixed up to space rotations $\SO(3)$.

For the massless electromagnetic vector fields the Lagrangian contains in addition to
the coupling constant $ e^2$ also the gauge fixing terms with a scalar field
$L(x)$ and gauge fixing parameter $\ka$
\begin{eqn}{l}
\bl L( A, F_A, L)=
 {1\over2}F_A^{jk}\ep^{lm}_{jk}\p_l A_m+L\p^jA_j
+ e^2  { F_A^{jk} F_{Ajk}\over4}+ \ka { L^2\over2}
\end{eqn}The curvature $F_A$ and the gauge fixing field $L$ 
are the canonical partners for the gauge field $A$.
The nontrivial commutators are given by
\begin{eqn}{rl}
&{\scriptsize\pmatrix{
\com{i F_A^{kl}}{ A^j}&\com { A^k}{ A^j}\cr
\com{ F_A^{kl}}{ F_A^{jm}}&\com { A^k}{i L}\cr}}(x)
={\scriptsize\pmatrix{
-i\ep_{ut}^{lk}\de^j_s{\p^u\over e^2   } &\de_t^k\de_s^j\cr
-\ep_{ut}^{lk}\ep_{rs}^{jm}{\p^r\p^u\over e^4  }
&-i\de^k_t{\p_s\over \ka  }\cr}}   \com{ A^t}{ A^s}(x)\cr
&~~\cr
&=
\int{d^4qe^{ixq}\over(2\pi)^3}\ep(q_0) \mu_e
{\scriptsize\pmatrix{
{q^u\over\mu_e }\ep^{kl}_{ut}\eta^{tj}\de(q^2)&
\ell_e^2[-\eta^{kj}\de(q^2)-{ e^2+\ka \over e^2}q^kq^j\de'(q^2)]\cr
{1\over\ell^2_e}\ep^{kl}_{ut}\ep^{jm}_{rs}\eta^{ts}{q^rq^u\over\mu_e^2}\de(q^2)
&{q^k\over \mu_e }\de(q^2)\cr}} \cr
\end{eqn}In the space-time translations analysis 
the vector field components are expressed in a Witt basis
\begin{eqn}{l}
\com{ A^k}{ A^j}(x)
=\int {d^3 q e^{-i\rvec x\rvec q}\over(2\pi)^3}{\mu_e\over q_0}
H({\rvec q\over q_0})_{\ul k}^k
~\brack{ A A}^{\ul k\ul j}(x_0) ~
H({\rvec q\over q_0})_{\ul j}^j\cr
\end{eqn}The transversal components (photons) have the $\SO(2)$-normalization
$\ell^2_e$,
the nondiagonal light fixing parameter $\al$ is combined from $ e^2$ and the gauge fixing
parameter $\ka $ 
\begin{eqn}{l}
\brack{ A A}(x_0)=\ell_e^2\Bigl\{ 
{\scriptsize\pmatrix{
0&0&-\al\cr 0&\bl 1_2&0\cr- \al&0&0\cr}}i\sin x_0q_0
+ix_0q_0\be{\scriptsize\pmatrix{
e^{-ix_0q_0}&0&0\cr 0&0&0\cr 0&0& e^{ix_0q_0}\cr}}   \} \cr
\ell^2_e={ e^2\over\mu_e}={\mu_e\over M^2},~~
\al=-{3  e^2+\ka \over 2 e^2},~~\be=-{ e^2+\ka \over 2 e^2}\cr
\end{eqn}
 
The transmutation to a Witt basis
is performed by $H({\rvec q\over q_0})$ 
(Sylvester indices $k,j,\dots=0,1,2,3$, 
Witt indices $\ul k,\ul j,\dots=0,1,2,3$)
\begin{eqn}{rl}
&H({\rvec q\over q_0})^k_{\ul k} =O({\rvec q\over q_0})_j^k\o w^j_{\ul k}
\cong{\scriptsize {1\over q^0}}{\scriptsize\pmatrix{
{q^0\over\sqrt2}&0&0&{q^0\over\sqrt2}\cr
{q^1\over\sqrt2}
&q^0-{(q^1)^2\over q^0+q^3}&{q^1q^2\over q^0+q^3}&-{q^1\over\sqrt2}\cr
{q^2\over\sqrt2}
&-{q^1q^2\over q^0+q^3}&q^0-{(q^2)^2\over q^0+q^3}&-{q^2\over\sqrt2}\cr
{q^3\over\sqrt2}&- q^1&- q^2&-{q^3\over\sqrt2}\cr}},~q^2=0  \cr
&H({\rvec q\over q_0}) 
{\scriptsize\pmatrix{0&0&-1\cr 0&\bl 1_2&0\cr- 1&0&0\cr}}
H({\rvec q\over q_0})=
{\scriptsize\pmatrix{-1&0\cr 0&\bl 1_3\cr}}\cr
\end{eqn}For the massless photons a rest system with
one space axis, e.g. a 3rd direction, has to be fixed up to  space plane
rotations $\SO(2)$.

The 1st and 2nd component have a particle structure with Fock value
\begin{eqn}{l}
\ul k,\ul j\in\{1,2\}:
\angle{\brace{ A A}}(x_0)\cong\ell_e^2 \bl1_2\cos x_0q_0
\end{eqn}

\chapter{Determination of\\ the Weinberg Angle}

In this chapter a symmetry normalization related framework is proposed
and conditions are given therein leading to a numerical
determination of the Weinberg angle.

To recapitulate:
The four coupling constants $g$
or masses $\mu$ in the electroweak triangle for the standard model
\begin{eqn}{rl}
\cl G=(g_1,g_2;g_Z|e)&
={1\over M}(\mu_1,\mu_W;\mu_Z|\mu_e)
=M(\ell_1^2,\ell_2^2;\ell_Z^2|\ell_e^2)\cr
& \sim({1\over 2.9},{1\over 1.6};{1\over 1.4}|{1\over 3.3})\cr
\end{eqn}contain the  
interaction symmetry normalizations $g_1^2$ for hypercharge $\U(1_2)$ and 
$g_2^2$ for isospin $\SU(2)$. The particle normalizations $\ell_Z^2$
and $\ell_e^2$ are related to the stability groups $\SO(3)$ and $\SO(2)$ resp.
of the neutral massive and massless particles 
resp. in the Lorentz group $\SO^+(1,3)$. $M^2$ normalizes the internal group 
$\U(2)$.

To compute the interaction and particle normalizations
(for a first  purely algebraic 
orientation without the interaction related running of the coupling 
constants), one can proceed in three steps:
First, 
the ratio of the $\U(1_2)$ and $\SU(2)$ normalizations has to be understood,
i.e. the Weinberg angle $\tan ^2\th={g_1^2\over g^2_2}$. 
An attempt for this first step is given in this chapter.
Therewith the
form of the electroweak triangle and the ratios of electroweak masses,
coupling constants and normalizations are determined. As a second step,
the absolute value of one internal coupling constant, e.g.
the fine structure constant ${e^2\over 4\pi}$, has to be given. 
Finally, the  $\U(2)$ unit $M^2$
has to be related to other masses,
e.g. to the Planck mass or to the proton mass.   

\section{Invariant Lie Algebra Forms}

Representations of Lie symmetries have their intrinsic metrical structure:
Any representation
of a Lie algebra $L$ on a vector space $V$, both finite dimensional, in
the endomorphisms algebra with its natural Lie algebra
structure
\begin{eqn}{l}
L\x V\map V,~~l\m v=\cl D(l)(v)\cr
\end{eqn}gives rise to the 
$V$-associated $L$-invariant multilinear trace forms
\begin{eqn}{l}
t_V^k:\underbrace{L\x\cdots\x L}_{k-\rm{times}}\map\K,~~
t _V^k(l_1,\dots,l_k)=\tr\cl D(l_1)\o\cdots\o \cl D(l_k)\cr
l\m t^k_V(l_1,\dots,l_k)=
t^k_V(\com l{l_1},\dots,l_k)+\dots+
t^k_V(l_1,\dots,\com l{l_k})=0\cr
\end{eqn}

Any  irreducible representation of a semisimple 
Lie algebra $L$ 
on a complex vector space $V$ gives an  invariant bilinear Lie algebra form,
unique up to a  normalization. These associated bilinear forms 
are multiples $\al t^2_V$, $\al\in\K$, of the 'double' trace
\begin{eqn}{l}
 t_V^2(~,~):L\x L\map\K,~~\cases{
 t_V^2(l,m)=\tr\cl D(l)\o\cl D(m)= t_V^2(m,l)\cr
 t_V^2(\com kl,m)+ t_V^2(l,\com km)=0\cr}
\end{eqn}The bilinear form  associated  to the adjoint representation is the
Killing form.

E.g. the real Lie algebras $\log\SU(n)$,
especially the isospin Lie algebra $\log\SU(2)$, 
can be spanned with the $(n^2-1)$ generalized
Pauli matrices 
as basis
\begin{eqn}{l}
\hbox{basis of $\log\SU(n)$: }\{l^a={i\over\sqrt2}\tau(n)^a\mid a=1,\dots,n^2-1\}\cr
\hbox{for }\SU(2):~~\{\tau(2)^a=\tau^a\mid\hbox{Pauli matrices}\}\cr
\hbox{for }\SU(3):~~\{\tau(3)^a=\la^a\mid\hbox{Gell-Mann matrices}\}\cr
\end{eqn}The generalized Pauli matrices  have the structure constants
\begin{eqn}{ll}
\tau(n)^a\tau(n)^b&={2\over n}\de^{ab}\bl 1_n+
(\de^{abc}+i\al^{abc})\tau(n)^c\cr
&\hbox{totally symmetric }\de^{abc}\in\R\cr
&\hbox{totally antisymmetric }\al^{abc}\in\R\cr
\end{eqn}The defining representation of $\log\SU(n)$ on $V\cong\C^n$
has as associated bilinear forms
\begin{eqn}{rl}
t^2_n(~,~):&\log\SU(n)\x\log\SU(n)\map\R\cr
&t^2_n(l^a,l^b)=\al_n\tr{i\over\sqrt2}\tau(n)^a{i\over\sqrt2}\tau(n)^b=
-\al_n\de^{ab},~~\al_n\in\R
\end{eqn}$\SU(n)$ has no  structure which distinguishes any
normalization $\al_n$.

Since for a semisimple Lie algebra the derived Lie algebra $\com LL$
coincides with $L$, 
all nontrivial $L$-representations have to be traceless, i.e.
there do not exist nontrivial invariant linear forms for
semisimple Lie algebras.

This is different for abelian Lie algebras where the 'single' trace is
a basis for the invariant linear forms
\begin{eqn}{l}
L\x V\map V,~~l\m v=\cl D(l)(v)\cr
t^1_V: L\map \K,~~\cases{t^1_V(l)=\tr\cl D(l)\cr t^1_V(\com kl)=0\cr}
\end{eqn}Therewith one obtains the invariant bilinear forms 
of abelian Lie algebras from the squared 'single' trace
\begin{eqn}{l}
t^{1,1}_V(~,~):L\x L\map\K,~~
t^{1,1}_V(l,m)=\tr\cl D(l)\tr \cl D(m)\cr
\end{eqn}

E.g. the real Lie algebras $\log\U(1_n)$ with basis 
$\{l^0={i\over\sqrt2}\bl 1_n\}$,
especially the hypercharge Lie algebra $\log\U(1_2)$, 
have as invariant bilinear forms in the defining complex $n$-dimensional 
representation
\begin{eqn}{rl}
t^{1,1}_n(~,~):&\log\U(1_n)\x\log\U(1_n)\map\R\cr
&t^{1,1}_n(l^0,l^0)=
\be_n(\tr {i\over\sqrt2}\bl 1_n)^2=-{n^2\over2}\be_n,~~\be_n\in\R
\end{eqn}Again, $\U(1_n)$ has no structure which distinguishes any
normalization $\be_n$.

\section{Relative Normalization\\of 
Hypercharge and Isospin} 

In the electroweak sector of the standard model
the internal group
comes as product group $\U(2)=\U(1_2)\o\SU(2)$, but apparently 
not as the direct product group $\U(1)\x \SU(2)$.
The correlation of hypercharge and isopin via the 
common subgroup $\U(1_2)\cap\SU(2)=\{\pm\bl 1_2\}$
can be  seen  on the colourless sector of the standard model\cite{RAIF,S921}
where fields with  integer and half integer isospin $T$
have always integer and
half integer hypercharge $Y$ resp., e.g. the left handed electron
fields $(T,Y)=({1\over2},-{1\over2})$, the right handed electron fields $(0,-1)$,
the $\SU(2)$ gauge fields $(1,0)$ etc.

The situation is more complicated for the colour nontrivial sector,
e.g. for the quark fields, where 
in addition to the  $\SU(2)$-center $\{\sqrt{\bl 1_2}=\pm\bl 1_2\}$
(two-ality) also
the $\SU(3)$-center
$\{\sqrt[3]{\bl 1_3}\}$ (tri-ality) has to be taken into account.
Triality is correlated to  third  integer hypercharges: 
E.g., the left handed quarks as
isospin 
doublets $d_T=2$ and colour triplets $d_C=3$ have hypercharge 
$Y={1\over d_Td_C}$.
This relation has been discussed to some extent in\cite{S926,S941}. 
I will consider
in this paper only the colour trivial electroweak sector.

The invariant bilinear forms for the defining representation of 
the real $n^2$-dimensional Lie algebra $\log\U(n)$ 
allow two real normalization constants $\be_n$ and $\al_n$ for
the Lie subalgebras $\log\U(1_n)$ (abelian) and $\log\SU(n)$ (simple) resp.
\begin{eqn}{l}
g_n(~,~):\log\U(n)\x\log\U(n)\map\R,~~g_n(l,m)\cases{
=t^{1,1}_n(l,m)+t^2_n(l,m)\cr
=\be_n\tr l\tr m+\al_n\tr l\o m\cr}
\end{eqn}These bilinear forms read in the basis with Pauli matrices
\begin{eqn}{l}
\hbox{basis of $\log\U(n)$: }\{l^0,l^a\}=
\{{i\over\sqrt2}\bl 1_n,{i\over\sqrt2}\tau(n)^a\}\cr
g_n(~,~)\cong -{\scriptsize\pmatrix{
{\be_n n^2+\al_n n\over2}&0\cr0&\al_n\bl 1_{n^2-1}\cr}}
\end{eqn}

The real 2-dimensional space of the invariant bilinear forms of $\log\U(n)$
(coefficients $\al_n$, $\be_n$)
can be decomposed into the bilinear 
forms with symmetric and antisymmetric 'double' trace,
called Fierz symmetrical and antisymmetrical forms with
normalizations $\al_n^\pm$
\begin{eqn}{l}
g^\pm_n(~,~):\log\U(n)\x\log\U(n)\map\R,~\cases{
g^\pm_n(l,m)=\al_n^\pm(\tr l\tr m\pm\tr l\o m)\cr
\al_n^\pm\in\R\cr }\cr
g^+_n(~,~)\cong -\al_n^+{\scriptsize\pmatrix{
{n+1\choose2}&0\cr0&\bl 1_{n^2-1}\cr}},~~
g^-_n(~,~)\cong -\al_n^-{\scriptsize\pmatrix{
{n-1\choose2}&0\cr0&-\bl 1_{n^2-1}\cr}}\cr
\end{eqn}Only the Fierz symmetrical forms $g_n^+$ are definite. 
The indefinite Fierz antisymmetrical forms have signature $(1,n^2-1)$.
The invariance groups of the quadratic
forms $g_n^+$ and $g_n^-$ are 
the compact group $\O(n^2)$ and the noncompact group 
$\O(1,n^2-1)$ with $\SO(n^2-1)$
 embedding $\U(n)/\U(1_n)$.

For the hyperisospin group $\U(2)$ one obtains
as Fierz (anti)symmetrical quadratic forms
\begin{eqn}{l}
g^+_2(~,~)\cong -\al_2^+{\scriptsize\pmatrix{
3&0\cr0&\bl 1_3\cr}},~~
g^-_2(~,~)\cong -\al_2^-{\scriptsize\pmatrix{
1&0\cr0&-\bl 1_3\cr}}\cr
\end{eqn}If one can give an argument in favour of the 
definite Fierz symmetrical form $g_2^+$
as the physically relevant hyperisospin form (next section), 
the ratio of the normalizations
of abelian $\U(1_n)$-symmetry and simple $\SU(n)$-symmetry is fixed
\begin{eqn}{l}
\U(2):g_2^+(~,~)\cong{\scriptsize \pmatrix{
{1\over g_1^2}&0\cr 
0&{1\over g_2^2}\bl 1_3\cr}}
\then \tan^2\th={g_1^2\over g_2^2}={1\over 3}\cr
\end{eqn}and in the general case 
\begin{eqn}{l}
\U(n):g_n^+(~,~)\cong{\scriptsize\pmatrix{
 {1\over g_1^2}&0\cr 
0&{1\over g_n^2}\bl 1_{n^2-1}\cr}}
\then \tan^2\th={g_1^2\over g_n^2}={1\over{n+1\choose 2}}\cr
\end{eqn}

\section{Fierz (Anti)Symmetrical Forms}

Statistical structures can determine the Fierz (anti)symmetry of
the bilinear forms for the $\U(2)$-normalization: The endomorphisms
of a finite dimensional vector space $V$ 
are isomorphic to the tensor product $V\ox V^T$ of space and dual space,
the invariant linear trace form  is the dual product
\begin{eqn}{l}
\tr: V\ox V^T\map\K,~~\cases{
\tr v\ox\om=\dprod\om v\cr
\tr\com fg=0\cr}\cr
\hbox{dual bases }\tr e^a\ox\d e_b=\dprod{\d e_b}{e^a}=\de_b^a\cr
\end{eqn}

The extension of the trace from $V\ox V^T$ as 'double trace' for 
invariant linear forms for the power two tensors $(V\ox V^T)^2$ 
can be done in two ways,  symmetrically  or antisymmetrically
\begin{eqn}{l}
\tr_\pm^2:(V\ox V)\ox (V\ox V)^T\map\K\cr
\tr_\pm^2(v_1\ox v_2)\ox (\om_1\ox\om_2)=
\dprod{\om_1}{v_1}\dprod{\om_2}{v_2}
\pm \dprod{\om_1}{v_2}\dprod{\om_2}{v_1}
\end{eqn}These two possibilities are
the appropriate nontrivial forms for the symmetrical 
$\od $ or
antisymmetrical $\and$ 'square' of the vector space $V$
\begin{eqn}{l}
\tr_+^2:(V\od V)\ox (V\od V)^T\map\K\cr
\tr_-^2:(V\and V)\ox (V\and V)^T\map\K\cr
\end{eqn}

The gauge interactions of the standard model act on vector spaces   $V$
with $\U(2)$ representations. Those vector spaces carry in the quantum
structure either  Fermi or  Bose statistics, characterized by
$\com vu=0$ or $\acom vu=0$ resp. for $v,u\in V$. Therefore only
the symmetrical product $V\od V$ or 
the antisymmetrical product  $V\and V$  is nontrivial,
e.g. for the Higgs vectors $\phi\in\H\cong\C^2$ with Bose statistic  
\begin{eqn}{l}
2\phi^A\and \phi^B=\com{\phi^A}{\phi^B}=0,~~
\hbox{nontrivial }2\phi^A\od\phi^B=\acom{\phi^A}{\phi^B}
\end{eqn}

In a quantum structure where the hyperisospin $\U(2)$-interactions get
their normalization via a vector space 
acted on only with the defining $\U(2)$ representation, 
the associated bilinear form
for the Lie algebra of $\U(2)$ has to be either Fierz symmetrical or
antisymmetrical. The definite forms $g_2^+$
require  symmetrical combinations of the underlying 2-dimensional  
$\U(2)$ representations.

Obviously only the normalization ratio 
(mixing angle) for
$\log\U(1_2)$ (hypercharge)  and  $\log \SU(2)$ 
as Lie subalgebras of $\log \U(2)$ is fixed
by such a statistical argument, not
the absolute normalizations.

\end{document}